\documentclass{ws-ijmpa}
\usepackage{times}
\usepackage{overcite}

\begin{document}

\markboth{R.~Wischnewski for the Baikal Collaboration}
         {The Baikal Neutrino Telescope -- Results and Plans}

\catchline{}{}{}{}{}

\title{THE BAIKAL NEUTRINO TELESCOPE -- RESULTS AND PLANS}

\author{RALF~WISCHNEWSKI}
\author{for the BAIKAL COLLABORATION \\[1.5mm]
V.~AYNUTDINOV,$^{1}$
V.~BALKANOV,$^{1}$
I.~BELOLAPTIKOV,$^{2}$
N.~BUDNEV,$^{3}$
L.~BEZRUKOV,$^{1}$
A.~CHENSKY,$^{3}$
D.~CHERNOV,$^{4}$
I.~DANILCHENKO,$^{1}$
ZH.-A.~DZHILKIBAEV,$^{1}$
G.~DOMOGATSKY,$^{1}$
A.~N.~DYACHOK,$^{3}$
S.~FIALKOVSKY,$^{5}$
O.~GAPONENKO,$^{1}$
O.~GRESS,$^{3}$
T.~GRESS,$^{3}$
K.~KAZAKOV,$^{3}$
A.~KLABUKOV,$^{1}$
A.~KLIMOV,$^{6}$
S.~KLIMUSHIN,$^{1}$
K.~KONISCHEV,$^{1}$
A.~KOSHECHKIN,$^{1}$
L.~KUZMICHEV,$^{4}$
V.~KULEPOV,$^{5}$
VY.~KUZNETZOV,$^{1}$
B.~LUBSANDORZHIEV,$^{1}$
S.~MIKHEYEV,$^{1}$
M.~MILENIN,$^{5}$
R.~MIRGAZOV,$^{3}$
E.~OSIPOVA,$^{4}$
A.~PAVLOV,$^{3}$
G.~PAN'KOV,$^{3}$
L.~PAN'KOV,$^{3}$
A.~PANFILOV,$^{1}$
YU.~PARFENOV,$^{3}$
E.~PLISKOVSKY,$^{2}$
P.~POKHIL,$^{1}$
V.~POLECSHUK,$^{1}$
E.~POPOVA,$^{4}$
V.~PROSIN,$^{4}$
M.~ROSANOV,$^{7}$
V.~RUBTZOV,$^{3}$
Y.~SEMENEY,$^{3}$
B.~SHAIBONOV,$^{1}$
CH.~SPIERING,$^{8}$
O.~STREICHER,$^{8}$
B.~TARASHANSKY,$^{3}$
R.~VASILIEV,$^{2}$
E.~VYATCHIN,$^{1}$
R.~WISCHNEWSKI,$^{8}$
I.~YASHIN$^{4}$
and
V.~ZHUKOV$^{1}$
}

\address{
$^1$Institute for Nuclear Research of Russian Academy of Sciences,
    Moscow, RU-117312 Russia                                          \\
$^2$Joint Institute for Nuclear Research, Dubna, RU-141980 Russia     \\
$^3$Irkutsk State University, Irkutsk, RU-664003 Russia               \\
$^4$Skobeltsyn Institute of Nuclear Physics of Moscow State University,
    Moscow, RU-119234 Russia                                          \\
$^5$Nizhni Novgorod State Technical University, Niznii Novgorod,
    RU-603600 Russia                                                  \\
$^6$Kurchatov Institute, Moscow, RU-123182  Russia                    \\
$^7$St.\ Petersburg State Marine University, St.\ Petersburg,
    RU-190008  Russia                                                 \\
$^8$Deutsches Elektronen-Synchrotron (DESY), Zeuthen, D-15738 Germany
}

\maketitle

\pub{Received (25 October 2005)}{Revised (12 July 2005)}

\begin{abstract}
New results from the Baikal neutrino telescope NT$200$, based on the first $5$ 
years of operation (1998--2003), are presented. We derive an all-flavor limit
on the diffuse flux of astrophysical neutrinos between $20$~TeV and $50$~PeV, 
extract an enlarged  sample of high energy muon neutrino events, and obtain 
limits on the flux of high energy atmospheric muons. In 2005, the upgraded 
telescope NT$200+$ will be commissioned: $3$ additional distant strings with 
only $12$ photo-multipliers each will rise the effective volume to $20$~Mton at 
$10$~PeV for this largest running neutrino telescope in the Northern hemisphere.

\end{abstract}

\section{The Detector NT$200$}

The Cherenkov neutrino telescope is located in Lake Baikal, Siberia, at a depth 
of $1.1$~km. The present stage of the Baikal telescope, NT$200$, was put into 
operation in April, 1998. It consists of $192$ optical modules (OMs), mounted on 
$8$ vertical strings, each with $24$ pairwise arranged OMs -- altogether forming 
a cylinder of $40$~m diameter and $72$~m height.\cite{APP1}

Each OM contains a $37$-cm diameter photo multiplier (PM). The two PMs of a pair 
define a {\it channel}. They are switched in coincidence in order to suppress 
background from bioluminescence and PM noise. A {\it trigger} is formed by the 
requirement of $\geq N$~{\it hits} (with {\it hit} referring to a channel) 
within $500$~ns. $N$ is typically set to $3$ or $4$. For each event, amplitude 
and time of all fired channels are digitized and sent to shore. A separate
{\it monopole trigger} searches for hit patterns characteristic for non-
relativistic, bright objects like GUT monopoles ($\beta \sim 10^{-5}-10^{-3}$).

Lake Baikal deep water has an absorption length of
$L_{\rm abs}(480~\mathrm{nm})=20\div24~\mathrm{m}$, a scattering length of 
$L_s=15\div70~\mathrm{m}$ and a strongly forward peaked scattering.

We present selected results from  the first five  years of NT$200$ operation 
(April 1998 -- February 2003; $1038$ livetime days). We focus on the atmospheric 
muon neutrino sample, the diffuse high energy neutrino flux and on high energy 
atmospheric muons. For results on Magnetic Monopoles, WIMPs and GRBs see 
Ref.~\refcite{PARIS04}. We describe the upgraded telescope NT$200+$, to be 
commissioned in spring 2005.

\section{Atmospheric Muon Neutrinos}

The signature of charged current muon neutrino events is a muon crossing the 
detector from below. Muon track reconstruction algorithms and background 
rejection have been described elsewhere.\cite{APP2} The analysis of the full 
$5$-years sample re-optimizes for higher signal passing rate, allowing for a 
contamination of $15-20\%$ fake events.

\begin{figure}[h!]
\centering{
\epsfig{figure=                       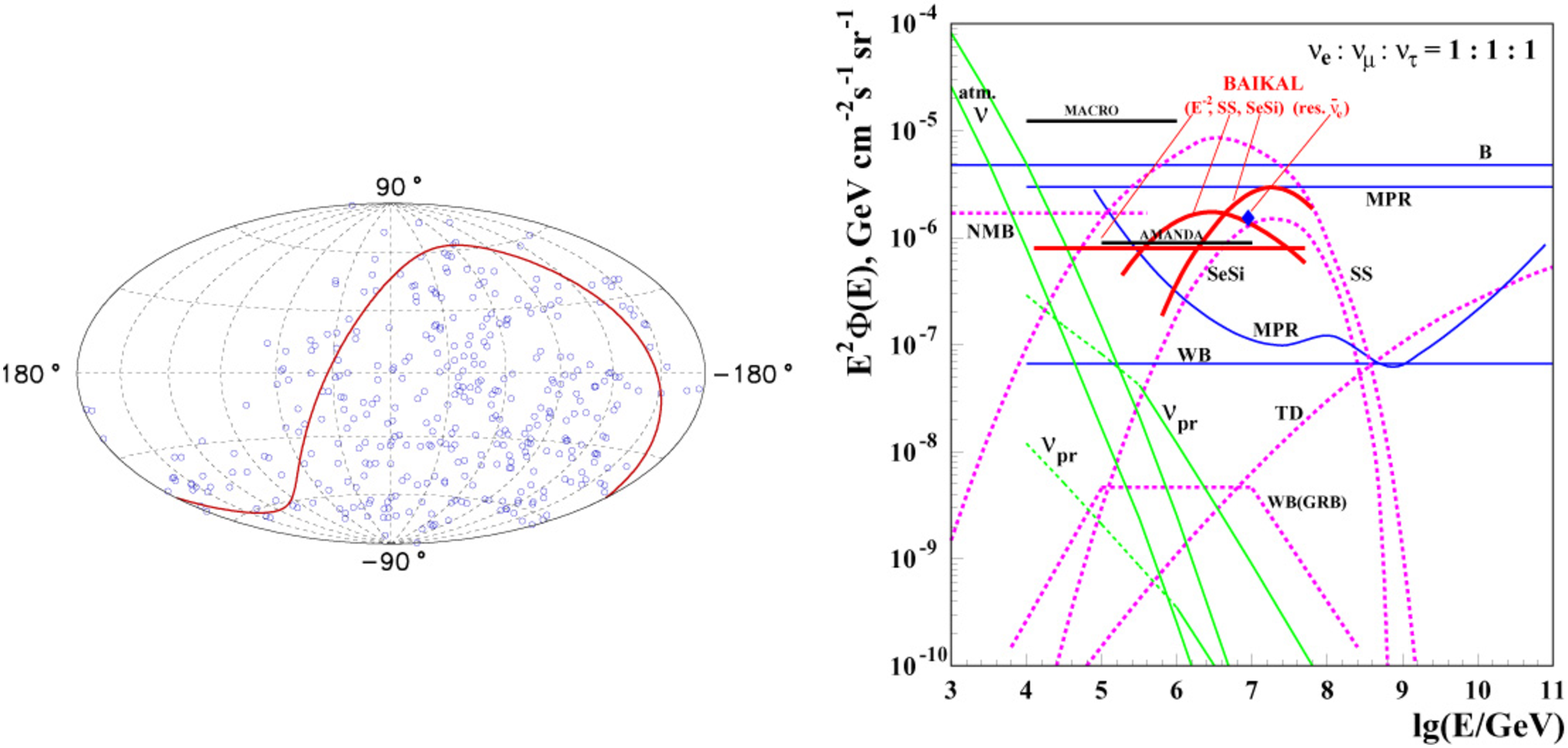,width=\linewidth}
}
\vspace*{8pt}
\begin{minipage}[t]{.47\linewidth}
\caption{Skyplot of neutrino events (galactic coordinates) for five years.}
\label{FIG_SKYPLOT}
\end{minipage}
\hfill
\begin{minipage}[t]{.47\linewidth}
\caption{Limits on diffuse neutrino flux from this analysis (Baikal), 
         compared to other experiments, theoretical bounds and  
         model predictions.}
\label{fig8}
\end{minipage}
\end{figure}

A total of $372$ upward going neutrino events are found. The arrival direction
distribution for this preliminary sample (skyplot in galactic coordinates)
is given in Fig.~\ref{FIG_SKYPLOT}.
Monte-Carlo simulations predict a total of $385$ atmospheric neutrino and 
background events, with an energy threshold of $15-20$~GeV.

\section{Search for Extraterrestrial High Energy Neutrinos}

The sensitivity of NT$200$ is maximal for $\nu_e$ and $\nu_{\tau}$ interactions 
(charged and neutral current) and $\nu_{\mu}$ neutral current interactions --
when bright electromagnetic and/or hadronic cascades are formed at the 
interaction vertex. For these ``cascade events'', light from the pointlike 
cascades  is visible from large distances. For vertices located far outside the 
telescope sufficient background rejection power is retained against the main 
background -- bright bremsstrahlung flashes along downgoing atmospheric muons 
passing far outside the array.\cite{APP3} Key point is the low level of light 
scattering in Baikal water, resulting in a sensitive volume significantly 
exceeding the geometric detector volume. This yields a sensitivity of NT$200$ 
comparable to the much larger AMANDA detector. 

The data analysis is essentially unchanged with respect to the detailed description 
for earlier data samples.\cite{PARIS04,APP3} It selects events with
(1) the minimum time difference of all combinations of hit channels along
    each string, $t_{\min}$, compatible with an upgoing plane light wave, 
    and 
(2) a large number of hit channels in the detector $N_{\rm hit}$
    (used as rough energy measure).
For $1038$ lifetime days, in total $3.45\times10^8$ events with 
$N_{\rm hit}\ge4$ have been recorded. The distribution of selected cascade 
events is compatible with the background from bright high energy 
muons.\cite{PARIS04,APP3} From the absence of any events in the signal region 
(i.e.\ with large  $N_{\rm hit}$), an upper limit on the diffuse flux of 
extraterrestrial neutrinos of a given spectrum is derived. The effective volume 
rises from $2\times10^5$~m$^3$ for $10$~TeV to $6\times10^6$~m$^3$ at $10^4$~TeV 
-- compared to a NT$200$ geometric volume of only $\approx10^5$~m$^3$.

For an $E^{-2}$ behaviour of the neutrino spectrum and a flavor ratio 
$\nu_e:\nu_{\mu}:\nu_{\tau}=1:1:1$ at earth,  the $90\%$ C.L.\ upper limit 
obtained in this analysis is:
\begin{equation}
E^2\Phi_{(\nu_e+\nu_{\mu}+\nu_{\tau})}<8.1\times 10^{-7}~%
\mbox{cm}^{-2}~\mbox{s}^{-1}~\mbox{sr}^{-1}~\mbox{GeV}.
\end{equation}
The model independent limit on $\bar{\nu}_e$ at the $W$-resonance energy is: 
\begin{equation}
\Phi_{\bar{\nu}_e}\leq3.3\times10^{-20}~%
\mbox{cm}^{-2}~\mbox{s}^{-1}~\mbox{sr}^{-1}~\mbox{GeV}^{-1}.
\end{equation}

Figure~\ref{fig8} shows our upper limits on ($\nu_e+\nu_{\mu}+\nu_{\tau}$) 
diffuse fluxes from AGNs shaped according to the models of Stecker and 
Salamon~(SS), of Semikoz and Sigl~(SeSi) and on an $E^{-2}$ spectrum according 
to Nellen {\it et~al}.~(NMB) (experiment -- solid, model -- dashed lines; for 
references to diffuse models asee Ref.~\refcite{PARIS04}).
The resonant $\bar{\nu}_e$ flux limit (2) is also shown (diamond). In addition, 
Fig.~\ref{fig8} gives experimental limits from MACRO\cite{MACROHE} and the 
AMANDA-II cascade search,\cite{AMANDAHE} theoretical bounds obtained by 
Berezinsky~(B), by Waxman and Bahcall~(WB), by Mannheim {\it et~al}.~(MPR), and 
predictions for neutrinos from topological defects (TD) and from GRB~(WB(GRB))
[see Ref.~\refcite{PARIS04} for the detailed references].

\section{Search for High Energy Muons}

Atmospheric muons and neutrinos are the most severe background source, when 
searching for extraterrestrial neutrinos. At GeV--TeV energy scale, atmospheric 
muons are predominantly produced in $\pi/K$-decays, with a well known steep 
power-law spectrum. Above the $10$~TeV scale, the predicted energy spectrum is 
subjected to large uncertainties: ``prompt'' muons with a harder spectrum and 
produced in semileptonic decays of charmed particles, are expected to dominate 
over the ``conventional'' $\pi/K$-muons. With heavy-quark production properties 
at these energies unknown, atmospheric muon flux measurements are needed.

\begin{figure}[h!]
\centerline{
\epsfig{figure=                       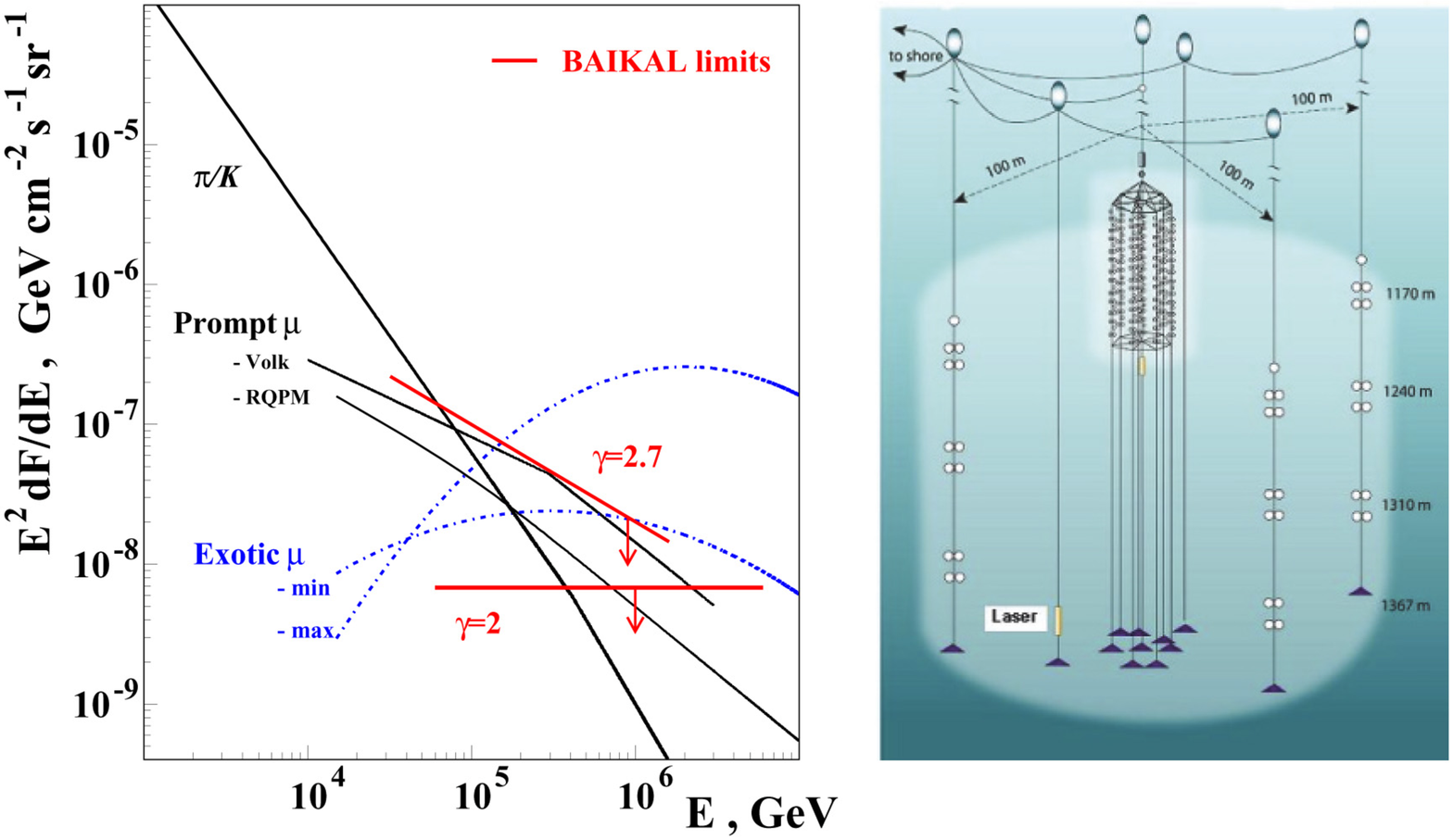,width=\linewidth}
}
\vspace*{8pt}
\begin{minipage}[t]{.54\linewidth}
\caption{Obtained upper limits on the high energy atmospheric muon flux 
         (curves with arrows for $\gamma=2$; $2.7$), and
         predicted fluxes (for $\pi/K$, prompt, exotic).}    
\label{fig9}
\end{minipage}
\hfill
\begin{minipage}[t]{.38\linewidth}
\caption{Sketch of the new NT$200+$ telescope: with NT$200$ at center, 
         and $3$ new outer strings at $100$~m radius.}
\label{FIG_NT200+}
\end{minipage}
\end{figure}

To search for high energy muons, we use the same event sample and cuts as for 
the diffuse high energy neutrino sample, to which atmospheric muons in turn are 
a background source (see above). With no events left at final cut level, upper 
flux limits can be derived for various muon energy spectra. The experimental 
limit for a typical prompt muon spectrum with $\gamma=2.7$, given in 
Fig.~\ref{fig9}, is still above theoretical 
predictions.\cite{PARIS04,VPROMPT,BUG}

We can also test for an ``exotic'' component of high energy atmospheric muons, 
postulated\cite{PETRUKHIN} to explain the ``knee'' in the cosmic ray energy 
spectrum by a new interaction at PeV-scale.\cite{HOERANDEL} The hard exotic 
spectra\cite{PETRUKHIN} (dashed curves in Fig.~\ref{fig9}) result in a high 
sensitivity of NT$200$. Our experimental limit\cite{HIGHEMU} for a generic
$E^{-2}$-spectrum is given in Fig.~\ref{fig9}. It shows, that the sensitivity is 
close to the lowest exotic flux. 

\section{The Upgraded Telescope NT$200+$}

To reach a significant sensitivity improvement for high energy cascade neutrino 
events, we are upgrading the Baikal telescope to the detector NT$200+$. This 
detector\cite{PARIS04} consists of NT$200$ and three additional external strings 
with $12$ PMs each at $100$~m radius from NT$200$, as sketched in 
Fig.~\ref{FIG_NT200+}. NT$200+$ has a $\nu_e$-effective  volume of $4$~Mton at 
$1$~PeV and of $20$~Mton at $100$~PeV neutrino energies, and allows for precise 
reconstruction of cascade vertex and energy. The sensitivity limit of NT$200+$ 
for a diffuse astrophysical flux of $\nu_e$'s, assuming $\gamma=2$ and a flavor 
ratio $\nu_e:\nu_{\mu}:\nu_{\tau}=1:1:1$ at earth, will be 
\[
E^2\Phi_{(\nu_e+\bar{\nu}_e)}<9\times10^{-8}~%
\mbox{cm}^{-2}~\mbox{s}^{-1}~\mbox{sr}^{-1}~\mbox{GeV}
\]
at $90\%$ C.L.\ and for three years of data.

The NT$200+$ will be commissioned in spring 2005; in 2004 two outer strings and 
the upgraded underwater data acquisition system had been installed. For the 
long-term future, a km$^3$-scale detector at Lake Baikal is under discussion: 
made of sparsely instrumented building blocks like NT$200+$ (with NT$200$ 
replaced by a single string).\cite{PARIS04}

\section{Conclusions}

The deep underwater neutrino telescope NT$200$ in Lake Baikal is taking data 
since 1998. For the full statistics (April 1998 -- February 2003), $372$ upward 
muon neutrino events are found. From the analysis of high energy cascade events 
we derive a limit on the diffuse flux of astrophysical neutrinos. We demonstrate 
the capability to set severe constraints for an exotic component in the 
atmospheric muon flux at PeV energies. In 2005, the upgraded detector NT$200+$ 
will start operation%
\footnote{At time of proof-reading, NT$200+$ had been fully commissioned.}
and allows to do astroparticle physics at the $10$~Mton scale with this largest 
high energy neutrino telescope in the Northern hemisphere.

\section*{Acknowledgements}

This work was supported by the Russian Ministry of Education and Science,
the German Ministry of Education and Research and the Russian Fund of Basic 
Research, grants $04-02-31006$, $02-02-17031$, $04-02-17289$ and
$02-07-90293$, Grant of President of Russia NSh-$1828.2003.2$ and by the
Russian Federal Program ``Integration'' (project no.~$248$).

\end{document}